\documentclass[pra,a4paper,superscriptaddress,twocolumn,amsmath,amssymb, longbibliography]{revtex4-1}
\pdfoutput=1
\usepackage[english]{babel}
\usepackage{graphicx}
\graphicspath{{./Figs/}}
\usepackage{color}
\usepackage{bm}
\usepackage{epstopdf}

\usepackage{verbatim}

\begin{document}
	

\title{A silicon-based single-electron interferometer coupled to a fermionic sea}
	
\author{Anasua Chatterjee}
\email{anasua.chatterjee.12@ucl.ac.uk}
\affiliation{London Centre for Nanotechnology, University College London, London, WC1H 0AH, United Kingdom}
\author{Sergey N.~Shevchenko}
\affiliation{Center for Emergent Matter Science, RIKEN, Wako-shi, Saitama 351-0198, Japan}
\affiliation{B. Verkin Institute for Low Temperature Physics and Engineering, Kharkov 61103, Ukraine}
\affiliation{V. Kazarin Kharkov National University, Kharkov 61022, Ukraine}
\author{Sylvain Barraud}
\affiliation{CEA/LETI-MINATEC, CEA-Grenoble, 38000 Grenoble, France}
\author{Rub\'en M. Otxoa}
\affiliation{Hitachi Cambridge Laboratory, J. J. Thomson Avenue, Cambridge CB3 0HE, United Kingdom}
\affiliation{Donostia International Physics Center, Donostia-San Sebastian 20018, Spain}
\affiliation{Department of Material Physics, Universidad del Pais Vasco, UPV/EHU, San Sebastian 20018, Spain}
\author{Franco Nori}
\affiliation{Center for Emergent Matter Science, RIKEN, Wako-shi, Saitama 351-0198, Japan}
\affiliation{Physics Department, University of Michigan, Ann Arbor, MI 48109-1040, USA}
\author{John J.~L.~Morton}
\affiliation{London Centre for Nanotechnology, University College London, London, WC1H 0AH, United Kingdom}
\affiliation{Department of Electronic \& Electrical Engineering, University College London, London WC1E 7JE, United Kingdom}
\author{M.~Fernando Gonzalez-Zalba}
\email{mg507@cam.ac.uk}
\affiliation{Hitachi Cambridge Laboratory, J. J. Thomson Avenue, Cambridge CB3 0HE, United Kingdom}	
	
\begin{abstract}
We study Landau-Zener-St{\"u}ckelberg-Majorana (LZSM) interferometry under the influence of projective readout using a charge qubit tunnel-coupled to a fermionic sea. This allows us to characterise the coherent charge qubit dynamics in the strong-driving regime. The device is realised within a silicon complementary metal-oxide-semiconductor (CMOS) transistor. We first read out the charge state of the system in a continuous non-demolition manner by measuring the dispersive response of a high-frequency electrical resonator coupled to the quantum system via the gate. By performing multiple fast passages around the qubit avoided crossing, we observe a multi-passage LZSM interferometry pattern. At larger driving amplitudes, a projective measurement to an even-parity charge state is realised, showing a strong enhancement of the dispersive readout signal. At even larger driving amplitudes, two projective measurements are realised within the coherent evolution resulting in the disappearance of the interference pattern. Our results demonstrate a way to increase the state readout signal of coherent quantum systems and replicate single-electron analogues of optical interferometry within a CMOS transistor.
\end{abstract}

\maketitle
\section{Introduction}

Silicon quantum electronics is a nascent field in which the discreteness of the electron charge or spin is exploited to obtain additional device functionalities beyond the capabilities of the current silicon microelectronics industry~\cite{Zwanenburg2013}. Some of the most promising outcomes of this research field include: single-electron devices~\cite{Fujiwara2001,Fuechsle2012} (performing logic operations at the device level~\cite{Mol2011a}, acting as spin filters for spintronic applications~\cite{Weber2014} and aiming to redefine the Ampere~\cite{Rossi2014}) as well as silicon-based quantum computers and memories based on the long spin coherence times that silicon can offer~\cite{Muhonen2014, Veldhorst2014, Veldhorst2015, Steger}.
A developing area of silicon quantum electronics is the application of the coherent quantum properties of single-electron charge states to realise electronic analogues to optical interferometry experiments~\cite{Dupont-Ferrier2013, Gonzalez-Zalba2016}. Optical interferometry has enabled the development of extremely sensitive detectors that, for example, have recently detected gravitational waves~\cite{Abbot2016}. However when electrons, instead of photons, are used for interferometry, it can lead to novel applications such as electron holography for precise imaging~\cite{Akashi2015} or testing the effect of Fermi-Dirac statistics in quantum optics~\cite{Bocquillon2013}.

\begin{figure*}
	\includegraphics[width=\textwidth]{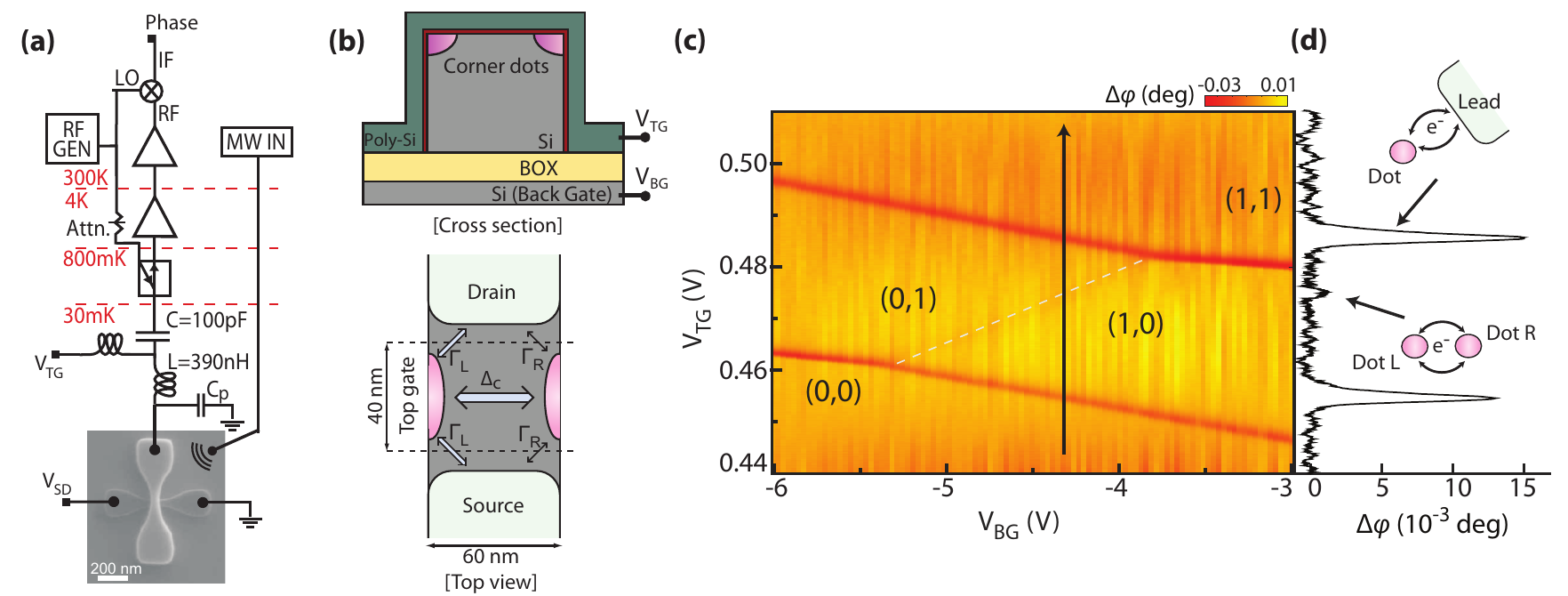}
	\caption{Dispersive detection of a DQD. (a) Scanning electron microscope image of a device similar to the one measured connected to a radio frequency reflectometry set-up via the top gate. An on-PCB coplanar waveguide close to the device is used to deliver variable-frequency MW signals. The device gate length is 40~nm and channel width is 60~nm. (b) Schematic of the device indicating the location of the corner quantum dots. The top panel represents a cross-section perpendicular to the direction of the current flow, where we can see the location of the dots. The back gate voltage ($V_\text{BG}$) is applied to the handle wafer. The bottom panel shows a top-view of the device in which the top gate appears transparent for clarity. Five different electronic transitions, between dots and reservoirs, are marked by arrows. $\Delta_\text{c}$ represents the tunnel coupling and $\Gamma_\text{L(R)}$ the relaxation rates between the left(right) dot and the reservoirs. (c) Resonator phase response $\Delta\varphi$ as a function of top gate and back gate voltages ($V_\text{TG}, V_\text{BG}$). The white dashed line emphasizes the interdot charge transition (ICT) and the black solid arrow indicates the position of the trace in panel (d). Electron numbers in the DQD appear in parenthesis. (d) $\Delta\varphi$ as a function of $V_\text{TG}$, showing the relative intensity of the ICT and the dot-to-sea transitions.}
	\label{fig1}
\end{figure*}

In quantum electronic devices, coupled two-level systems (TLS) can be used to coherently split single electron states via Landau-Zener (LZ) transitions~\cite{Oliver2005}. The quantum interference between consecutive LZ transitions gives rise to Landau-Zener-St{\"u}ckelberg-Majorana interferometry (LZSM) which encodes information about the system's coherent evolution and energy spectrum~\cite{Shevchenko2010}. This technique has been successfully applied to coherently control a variety of solid-state platforms such as superconducting qubits~\cite{Sillanpaa2006, Wilson2010}, charge and spin qubits in semiconductor quantum dots and dopants~\cite{Petta2010,Stehlik2012, Forster2014,Dupont-Ferrier2013}, and nitrogen vacancy centres in diamond~\cite{Zhou2014}, and has been used to address fundamental phenomena such as second-order phase transitions~\cite{Xu14, Gong16}. Although LZSM interferometry is typically described in terms of TLS, it has been suggested that more complex multi-level systems can be studied using LZSM interferometry. Such approach has been harnessed in multi-level superconducting qubits for high-resolution excited state spectroscopy~\cite{Berns2008} and in multi-level semiconductor qubits for extreme harmonic generation~\cite{Stehlik2014,Stehlik2016}.

In this Article, we present a LZSM interferometry study performed in a single-electron double quantum dot (DQD) formed at the edge states of a silicon transistor fabricated using industry-standard 300 mm silicon-on-insulator technology~\cite{Voisin2014}. The DQD is operated in the charge qubit regime and it is further coupled to a fermionic sea generating a multi-level energy spectrum. We read out the state of the charge qubit dispersively by interfacing the quantum system with a high-frequency electrical resonator, via the gate~\cite{Colless2013, Gonzalez-Zalba2015}. We demonstrate that, by tuning the amplitude of the driving microwave signal, we can access multiple LZSM regimes by introducing increasing degrees of projective readout arising from the interaction of the qubit with the fermionic reservoirs of the device, which first enhance the interferometric signal, and then suppress it completely. Finally, we develop a theoretical model to account for the qubit-resonator interaction that accurately describes the interference patterns and the dispersive signal enhancement.  We do so by combining the adiabatic-impulse model description of the LZSM problem with the device multi-level-dependent capacitance. Our results motivate further studies exploring electron quantum optics in silicon and using the enhancement of the qubit readout signal for high-fidelity state readout.

\section{Device and resonator}

We perform the experiment on a fully-depleted silicon-on-insulator nanowire transistor similar to the one seen in the scanning electron micrograph in Fig.\ref{fig1}(a). The nanowire height and width are 11~nm and 60~nm respectively, and it is covered by a 40~nm long wrap-around top-gate that can be biased ($V_\text{TG}$) to create an electron accumulation layer in the channel. Additionally, the device can be back-gated by applying a voltage ($V_\text{BG}$) to the silicon handle wafer. The full fabrication details can be found elsewhere~\cite{Betz2014}. In square-section nanowire transistors, electron accumulation happens first at the top-most corners of the transistor creating a double quantum dot (DQD) in parallel with the source and drain ohmic contacts as can be seen in the schematics in Fig.\ref{fig1}(b)~\cite{Voisin2014,Betz2016}. Measurements are performed at the base temperature of a dilution refrigerator (35~mK) using gate-based radio-frequency reflectometry as in Ref.~\cite{Gonzalez-Zalba2016}. For this purpose, we embed the device in a tank circuit composed of a surface mount inductor, $L=390$~nH, and the parasitic capacitance to ground of the device, $C_\text{p}=$~660~fF, and perform homodyne detection at the resonant frequency of the resonator $f_\text{rf}=313$~MHz, see Fig.\ref{fig1}(a).

The demodulated phase response of the resonator ($\Delta\varphi$) is sensitive to regions of charge instability in single-electron devices and more particularly to parametric capacitance changes, $C_\text{pm}$, that occur when electrons tunnel~\cite{Mizuta2017,Shevchenko12,Shevchenko15,LaHaye09,Okazaki16}. The two variables are related by $\Delta\varphi=-\pi Q C_\text{pm}/C_\text{p}$, where $Q$ is the loaded Q-factor of the resonator. We use this feature of our device-resonator system to measure dispersively the charge stability diagram of the device as a function of $V_\text{TG}$ and $V_\text{BG}$, as shown in Fig.\ref{fig1}(c). In the subthreshold regime of the transistor, where the source-drain current is too low to be measured, we observe the characteristic signature of a DQD in the few-electron regime. We identify four different stable charge configurations indicated by ($nm$) where $n$ and $m$ indicate the number of electrons in the dots. Since we do not observe any additional charge transition at lower gate voltages, we conclude that at low gate voltages the system is depleted of electrons. Hence, our device is operated in the single-electron charge qubit regime where an electron can occupy the left or right dot --- states (10) and (01), respectively --- and can also unload or load an electron via interaction with the fermionic seas of the highly doped source and drain --- states (00) and (11), respectively.

The parametric capacitance of a DQD as seen from the top-gate is:
\begin{equation}\label{Cdiff}
	C_\text{pm}\approx -e\frac{\partial}{\partial V_\text{TG}}\left\{\alpha_\text{L}\left\langle n_\text{L}\right\rangle+\alpha_\text{R}\left\langle n_\text{R}\right\rangle\right\},
\end{equation}

\noindent where $\alpha_\text{L(R)}$ represents the left (right) top-gate coupling 
and $\left\langle n_\text{L(R)}\right\rangle$ is the average electron occupation of the left (right) dot~\cite{Gonzalez-Zalba2016}. To account for the interaction with the DQD with the fermionic reservoirs, we express this capacitance in terms of the occupation probabilities of the four charge states ($P_{nm}$)~\cite{Gonzalez-Zalba2016}: 
\begin{equation}\label{cap4}
	C_\text{pm}\approx 2e^2\alpha_-^2\frac{\partial}{\partial\varepsilon_\text{0}}\left\{P_{01}-P_{10}+\frac{\alpha_+}{\alpha_-}(P_{00}-P_{11})\right\}.
\end{equation}

Here, we have introduced the notation $\alpha_\pm=(\alpha_\text{L}\pm\alpha_\text{R})/2$ and the energy detuning between dots $\varepsilon_0=-2e\alpha_-(V_\text{TG}-V_\text{TG0})$, where $V_\text{TG0}$ is the top-gate voltage at which the (10) and (01) states hybridize (this value depends on $V_\text{BG}$). Intuitively, and as illustrated in Eq.~(\ref{cap4}), in the common situation where the two dots have similar coupling to the top gate (i.e.\ $\alpha_-$ is small), then states 01 and 10 are harder to distinguish through RF reflectometry. In contrast, transitions to the (00) and (11) states yield a larger change in capacitance, characterised by the  geometric factor $\alpha_+/\alpha_-=18$ in our device, measured using magnetospectroscopy~\cite{Sellier2007} (see Supplementary Information~\cite{Supp}).

The difference in dispersive response between states involving the interdot charge transitions (ICT) and dot-to-sea transitions (DST) can be  observed in Fig.~\ref{fig1}(d), where we plot $\Delta\varphi$ against $V_\text{TG}$. The DST transitions are well resolved; however, the ICT (indicated by the white dashed line in Fig.~\ref{fig1}(c)) is particularly faint due to the similar gate-couplings in our DQD, with a signal intensity ratio of $\sim15$, in agreement with the measured ratio $\alpha_+/\alpha_-$. Overall, these results demonstrate that transitions that involve the fermionic reservoirs have a larger dispersive signal with respect to the ICTs and the intensity ratio is determined by the geometrical factor $\alpha_+/\alpha_-$. 

\section{LZSM interferometry}

\begin{figure}
	\includegraphics[width=\columnwidth]{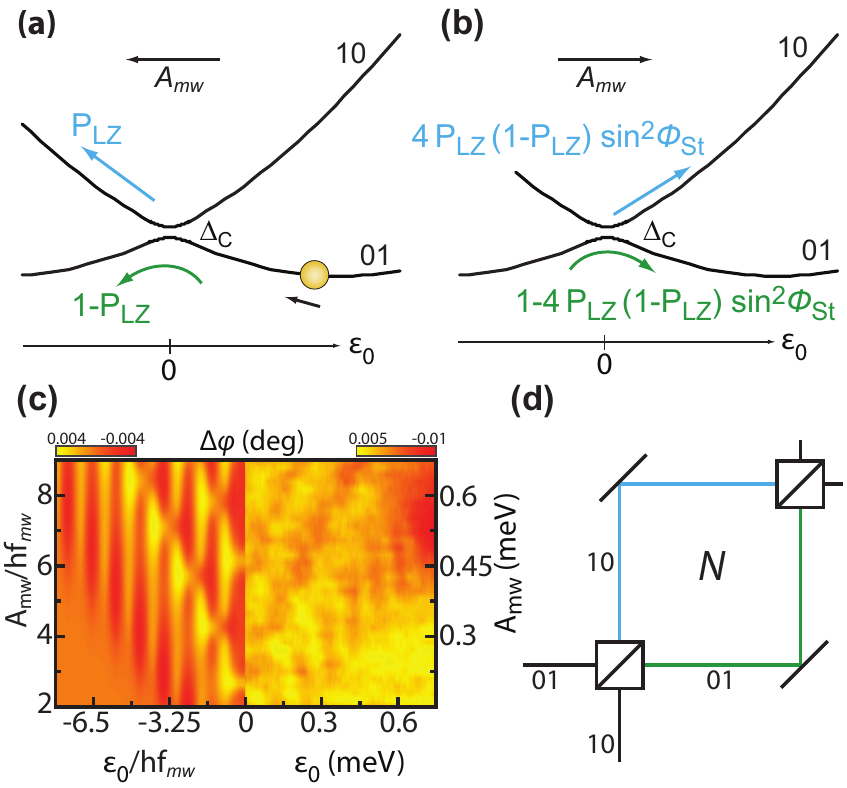}
	\caption{
Multi-passage LZSM interferometry. (a) Schematic of the first Landau-Zener transition. Starting from a well-defined state (01) a rapid passage through the anticrossing splits the electron wavefunction into the (01) and (10) states with probabilities $P_\text{LZ}$ and $1-P_\text{LZ}$ respectively. (b) Schematic of the probability distribution after the second passage. (c) $\Delta\varphi$ calculated (left of panel) vs experimental (right of panel). For the calculation we use $T_1=1.25$~ns, $T_2=0.25~$ns and $\Delta_\text{c}=34~\mu$eV. Here the photon energy is $hf_\text{mw}=87~\mu$eV. (d) Optical interferometry analogue showing the photon paths (electron charge states), the beam splitters (fast passage through the anticrossing) and refocusing mirrors (microwave electric fields). Here the process is repeated $N$ times. }
	\label{fig:multipass}
\end{figure}

Coherent LZSM interference occurs when a system is repeatedly (at least twice) driven through an anticrossing (of energy $\Delta_\text{c}$) at a rate comparable to $(\Delta_\text{c}/h)^2$, and over timescales shorter than the coherence time $T_2$.
For example, consider the situation in the vicinity of the ICT where a single electron can be driven back and forth between the two QDs, schematically represented in Fig.~\ref{fig:multipass}(a,b), with the anticrossing centered (by definition) at $\varepsilon_0=0$. We assume the system begins in state (01) (i.e. $\varepsilon_0>0$), and then $\varepsilon_0$ is driven to a negative value so the system traverses the anticrossing between the (01) and (10) states. For suitable sweep rates (neither sufficiently slow to be in the adiabatic limit, nor sufficiently fast to remain in the (01) state) this splits the electron wavefunction into two components which then acquire different dynamical phases. If $\varepsilon_0$ is swept back to a positive voltage, such that the system undergoes a second passage, the phase difference acquired by the two charge states leads to an interference pattern in the resulting probabilities of the (01) and (10) states.

\subsection{Multi-passage regime}

In our experiment, we produce a harmonic driving electric field by applying a microwave signal (of amplitude $A_\text{mw}$ and frequency $f_\text{mw}=21$~GHz) to an on-PCB antenna that effectively varies $V_\text{TG}$ at a fixed $V_\text{BG}$ as indicated by the black line in Fig.~\ref{fig1}(c). By changing the amplitude of the drive $A_\text{mw}$, or the starting position ($\varepsilon_0$, and thus $V_\text{TG}$) the phase acquired during the passages changes, resulting in interference fringes in the measured RF reflectometry signal Fig.~\ref{fig:multipass}(c).

To model the results we use a full unitary description of the each anti-crossing passage and dynamical phase acquired (see Supplementary Information~\cite{Supp}); for conciseness we summarise the reasoning using the probability, $P_\text{LZ}$, that an electron performs an LZ transition to the excited state following a passage:
\begin{equation}
	P_\text{LZ}=\text{exp}\left(-\frac{\pi\Delta_c^2}{2hf_\text{mw}\sqrt{A_\text{mw}^2-\varepsilon_0^2}}\right).
\end{equation}
Assuming the system starts in state (01), after two passages, the probability of the system returning to (01) is:
\begin{equation}
	P_\text{LZ,2}=1-4P_\text{LZ}(1-P_\text{LZ})\text{sin}^2\phi_\text{St}, \label{double}
\end{equation}

\noindent where we can see the interference term containing the St{\"u}ckelberg phase $\phi_\text{St}=\phi_\text{St}(\varepsilon_0,f_\text{mw},A_\text{mw}$) that captures the phase difference acquired during the free evolution. If the charge coherence is preserved for even longer timescales, the system can perform multiple correlated passages leading to a stationary probability distribution in $P_{01}$, given by

\begin{align}
P_\text{LZ,N}&=\frac{1}{2}\left[1+\text{sgn}(\varepsilon_0)(1-2P_\text{LZ,N}^+)\right], \\ \label{multiple}
P_\text{LZ,N}^+&=\frac{1}{2}\sum_k\frac{\Delta _{\text{c},k}^{2}}{\Delta
_{\text{c},k}^{2}+\frac{T_{2}}{T_{1}}(\left\vert \varepsilon _{0}\right\vert -khf_\text{mw})^{2}+\frac{\hbar ^{2}}{T_{1}T_{2}}}, 
\end{align}

\noindent where $\Delta _{\text{c},k}=\Delta_\text{c}J_k(A_\text{mw}/hf_\text{mw})$, with $J_k$\ standing for the Bessel function of the $k^{th}$ order and $T_1$ corresponds to the charge relaxation time. This probability can be converted to a capacitance using Eq.~(\ref{cap4}) and assuming that $A_\text{mw}$ is not large enough to reach the crossings with the (00) and (11) states ($P_{00}=P_{11}=0$):

\begin{equation}
C_\text{pm}\simeq 4e^{2}\alpha _{\mathrm{-}}^{2}\frac{\partial}{\partial\varepsilon _{0}}P_\text{01}\quad\text{and}\quad P_\text{01}=P_\text{LZ,N}.  \label{CQ_TLS}
\end{equation}

Both data and model in Fig.~\ref{fig:multipass}(c) show the characteristic signatures of multi-passage LZSM: enhanced $\Delta\varphi$ at equally-spaced points in $\varepsilon_0$, separated by the photon energy $hf_\text{mw}$, and (quasi)periodic $\Delta\varphi$ oscillations for increasing $A_\text{mw}$. We note that the interference patterns disappear for $f_\text{mw}<4$~GHz indicating a charge coherence time $T_2\sim0.25$~ns (see~\cite{Supp}). Since we are dealing with a classical resonator ($f_\text{rf}\ll k_\text{B}T/h, f_\text{mw}$), the resonator sees a stationary value for the occupation probabilities so that Eq.~(\ref{CQ_TLS}) holds. In our simulation, we use $\Delta_\text{c}=34~\mu$eV (extracted from the FWHM of the ICT (Fig.1(d))), $T_2=0.25$~ns, as measured experimentally, and find the best fitting $T_1$. We find the best match between the calculation and experiment for $T_1=5T_2=1.25$~ns.

The electron manipulation in this section resembles a multi-passage Mach-Zehnder interferometer, see Fig.~\ref{fig:multipass}(d). Here, the role of the beam splitter is played by the anticrossing that splits the electronic wavefunction into a superposition of two charge states~\cite{Petta2010} and the role of the phase difference is played by $\phi_\text{St}$. The microwave drive refocuses the different electron paths towards the anticrossing and the process is repeated $N=T_2f_\text{mw}\approx 5$ times. Overall, these results demonstrate that, although we continuously monitor the state of the qubit, the non-demolition nature of gate-based readout does not prevent the qubit from performing multiple coherent passages through the anticrossing.

\section{LZSM Interferometry with Projective readout}

\begin{figure*}
	\includegraphics[width=\textwidth]{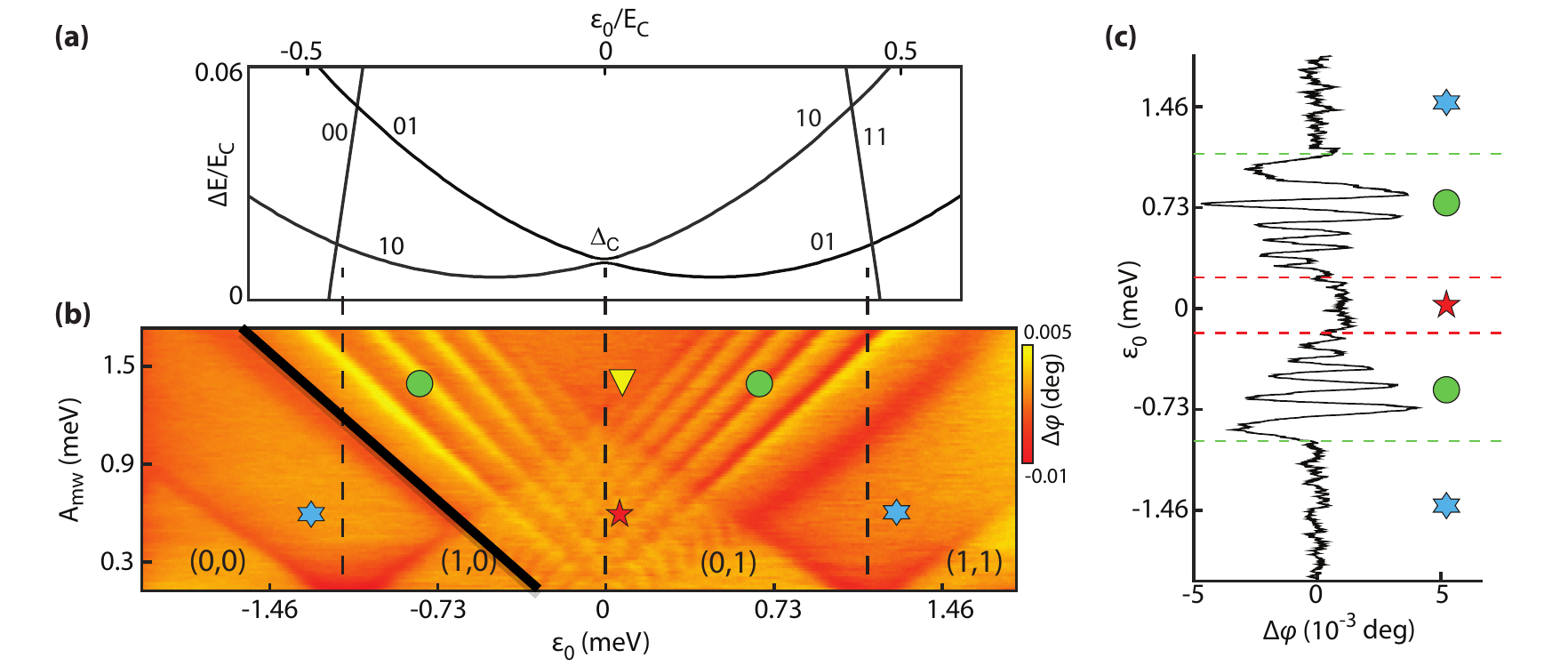}
	\caption{LZSM interferometry with projective readout. (a) Calculated energy level diagram of the DQD as a function of reduced detuning $\varepsilon_0/E_\text{C}$, for $\alpha_-$=0.05 and $E_\text{m}/E_\text{C}=10$. States (10) and (01) anticross at $\varepsilon_0=0$ and these cross the (00) and (11) states at larger $\left|\varepsilon_0\right|$. The small asymmetry between the dots places the two crossings with the (00) states close in $\varepsilon_0$ space as is also the case for the crossings with the (11) state. (b) $\Delta\varphi$ as a function of the detuning, $\varepsilon_0$, and amplitude of the MW, $A_\text{mw}$, for $f_\text{mw}=21$~GHz. The regions of constant $\Delta\varphi$, where electron numbers are well defined, are indicated by the numbers in parenthesis. The position of the (10)-(01) anticrossing and the (00)-(10) and (01)-(11) crossings are indicated by the black dashed lines. The incoherent LZSM, multi-passage LZSM, double-passage LZSM and single-passage regions are indicated by the blue star, red star, green circle and yellow triangle respectively. (c) $\Delta\varphi$ vs $\varepsilon_0$ trace at $A_\text{mw}= 0.55$~meV and $f_\text{mw}=11$~GHz. Symbols as in panel (b).}
	\label{fig:lzsm_all}
\end{figure*}

We next consider the result of larger driving fields ($A_\text{mw}$) or detunings ($\varepsilon _{0}$) for which transitions to the (00) and (11) states come into play, with two significant impacts. First, as these transitions involve charger transfer into a fermionic sea, they are not phase-preserving, as we shall see later, and hence have the effect of introducing projective readout of the charge. Second, as described above, transitions involving the fermionic sea yield a much larger measured reflectometry signal $\Delta\varphi$, than that associated to the qubit states (01) and (10).

To understand the coherent evolution through the multiple charge configurations, in Fig.~\ref{fig:lzsm_all}(a), we plot the full calculated energy level spectrum of the DQD as a function of reduced detuning $\varepsilon_0/E_\text{C}$, for a DQD with a top-gate coupling asymmetry $\alpha_-=0.05$, as in the case of our device. Here, $E_\text{C}$ is the charging energy of the QDs (see~\cite{Supp}). Additionally, in Fig.~\ref{fig:lzsm_all}(b), we plot a measurement of $\Delta\varphi$ as a function of $\varepsilon_0$ and microwave amplitude $A_\text{mw}$ across a range which enables the full exploration LZSM around the four charge states of the DQD. For small $A_\text{mw}$ and far from the charge transitions, the charge state is stable under microwave driving (such that $\Delta\varphi$ remains constant) and these regions are labeled accordingly in Figure~\ref{fig:lzsm_all}(b).

Elsewhere, the system evolution involves charge states transitions, leading, in some cases, to interferometric patterns in $\Delta\varphi$. We identify three distinct LZSM regimes involving the ICT anticrossing, indicated by the different symbols in Fig.~\ref{fig:lzsm_all}(b). First, for small $A_\text{mw}$ and $\varepsilon _{0}$, there is the region of multi-passage LZSM interference already described above (red star). As $A_\text{mw}$ and $\varepsilon _{0}$ are increased, the DSTs are crossed, producing a double-passage LZSM interference region where projective readout via the (00) and (11) states is performed every second passage (green circle). As a result of the DST, the amplitude of the interference pattern in this regime is as much as 8 times greater in the double-passage regions compared to the multi-passage region involving only the ICT, and this is further illustrated in Fig.~\ref{fig:lzsm_all}(c), where we plot $\Delta\varphi$ against $\varepsilon_0$ for $A_\text{mw}=0.55$~meV and $f_\text{mw}=11$~GHz. For yet larger values of $A_\text{mw}$, and while $\varepsilon _{0}$ remains small, the systems enters a regime where the projective readout via the (00) and (11) states occurs after every passage (single-passage regime, yellow triangle), prohibiting the manifestation of any interference effects. Finally, we highlight in the figure, regions of large $\varepsilon _{0}$ and small $A_\text{mw}$ which involve only the incoherent DSTs (and not the ICT), in which any interference signature is also absent. These regions are indicated in Fig.~\ref{fig:lzsm_all}(b) by blue stars. In the following sections, we explore each of these regimes in further detail.

\subsection{Incoherent regime of dot-sea transitions}

For small $A_\text{mw}$ and large $\varepsilon _{0}$, only the DSTs ((00)$\leftrightarrow$(10) or (01)$\leftrightarrow$(11)) are involved. Note that here only the left QD exchanges electrons cyclically with the fermionic sea of the source and drain reservoirs. The dynamics of such TLS have been described elsewhere~\cite{Persson2010, Gonzalez-Zalba2015} showing that when the tunnel-coupling to the reservoir is small compared to the driving rate, the system performs LZ transitions with probability close to unity. However, due to the large degeneracy of the electronic states in the reservoirs, relaxation times are often short and hence passages across the DST therefore yield no LZSM interference, but instead have the function of projective readout in the dynamics described below. This lack of interference pattern is captured in our data as can be seen in Fig.~\ref{fig:lzsm_all}(b). The transition between this regime and the regions of stable charge states enable us to calibrate the amplitude of the microwave electric field on the device, by relating it to $\varepsilon_0$ (see~\cite{Supp}). 

\subsection{Double-passage regime}

\begin{figure*}
\includegraphics[width=\textwidth]{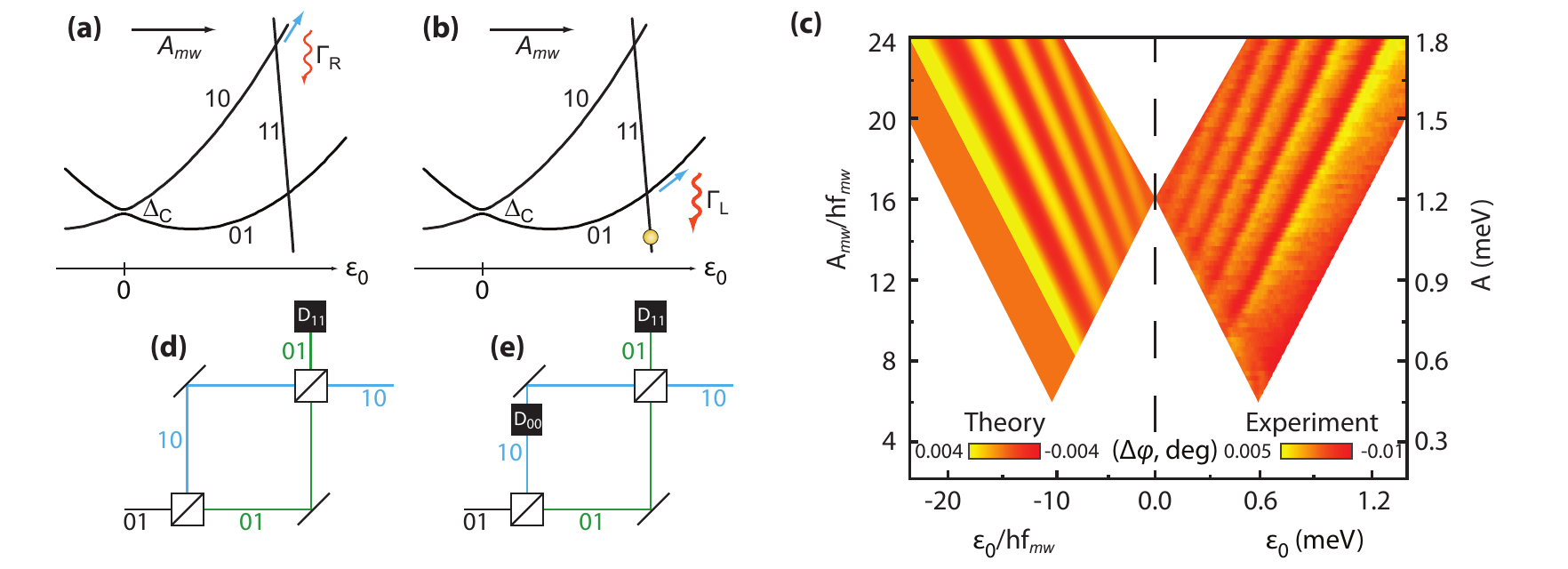}
\caption{Double-passage LZSM interferometry. (a) Schematic of the Landau-Zener transition at the (10)-(11) crossing after the double passage. The blue arrow indicates $P_\text{LZ}\approx 1$ and the faint red arrow indicates that the relaxation is slow after this crossing. (b) Schematic of the Landau-Zener transition at the (01)-(11) crossing after (a). Again $P_\text{LZ}\approx 1$ (blue arrow) but fast relaxation projects the system to the (11) state. (c) $\Delta\varphi$ calculated (left half of the panel) versus experimental (right half of the panel). For the calculation we use $\Delta_\text{c}=34~\mu$eV and $hf_\text{mw}=87~\mu$eV, for $\varepsilon_0/A_\text{mw}<1$. (d) Optical interferometry analogue now including the detector D$_{11}$ after the second passage [fast relaxation at the (01)-(11) transition]. (e) Optical interferometry analogue for the single-passage remine including an additional detector D$_{00}$ after the first passage [relaxation at the (10)-(00) transition].}
	\label{fig:doublepass}
\end{figure*}

We move on now to the double-passage regime indicated by the green circles in Fig.~\ref{fig:lzsm_all}(b). Following a double passage over the anticrossing, driven by the microwave electric field, the expected probability distribution of the (01) and (10) states is given by Eq.~(\ref{double})). At this point, due to the larger microwave amplitude, the system approaches a DST in which both (01) and (10) states cross (e.g.) the (11) state, as in Fig.~\ref{fig:doublepass}(a,b).

In order to understand the charge dynamics at this point, we require insights into the relaxation rates between the dots and the reservoirs. We can estimate the relaxation rate from the right dot to the reservoirs, $\Gamma_\text{R}$, from a measurement of the temperature dependence of the FWHM of the (10)$\leftrightarrow$(11) DST (taken at $V_\text{BG}=-3$~V as in Fig.~\ref{fig1}(c)). We obtain $\Gamma_\text{R}<12$~GHz (see~\cite{Supp}). Moreover, by analyzing the decay of the amplitude of the $\Delta\varphi$ oscillations towards low $\varepsilon_0$ (as in Fig.~\ref{fig:lzsm_all}(c)), we can extract a relaxation rate for the left dot $\Gamma_\text{L}\approx 50$~GHz (See~\cite{Supp}). We find a relaxation rate ratio $\Gamma_\text{R}/\Gamma_\text{L}<0.25$, indicating that the relaxation rate from the left dot to the reservoir is much faster than the relaxation from the right dot. Such an asymmetry in the relaxation rates, combined with the small difference in $\varepsilon_0$ between the (10)$\leftrightarrow$(11) and (01)$\leftrightarrow$(11) DSTs, results in relaxation occurring primarily via the left dot after the double-passage. This projects the system into the (11) or (00) state and then subsequent passages through the anticrossing are uncorrelated.

To confirm this description of the system dynamics, we present the experimental results, in Fig.~\ref{fig:doublepass}(c), alongside a simulation based on Eq.~(\ref{cap4}). Although in this regime $P_{01},P_{10},P_{11}\neq 0$, the main contribution to the capacitance will be from the $P_{11}$ term due to the large factor $\alpha_+/\alpha_-$

\begin{equation}
C_\text{pm} \approx 2e^2\alpha_-\alpha_+\frac{\partial}{\partial\varepsilon _{0}}P_{11}.
\label{CQ_enhanced}
\end{equation}%

In the limit where $A_\text{mw}\gg hf_\text{mw}$ and $\varepsilon_0$ is close to the (01)$\leftrightarrow$(11) crossing point, Eq.~(\ref{CQ_enhanced}) takes the form

\begin{equation}
C_\text{pm} \approx 2e^2\alpha_-\alpha_+\frac{\Gamma_\text{L}-\Gamma_\text{R}}{2f_\text{mw}}\frac{\partial}{\partial\varepsilon _{0}}P_\text{LZ,2}.
\label{CQ_enhanced2}
\end{equation}%

Both the data and simulation show (quasi)periodic oscillations in $\Delta\varphi$ for increasing $A_\text{mw}$ as in the multiple-passage regime. However, in contrast to the multi-passage LZSM, the periodic enhancement of $\Delta\varphi$ at regular values of $\varepsilon_0$ is now absent because only two consecutive passages are correlated~\cite{Berns2006}. The calculated results in the left part of the panel are obtained with Eq.~(\ref{CQ_enhanced2}) using $\Delta_\text{c}=34~\mu$eV, $f_\text{mw}=21$~GHz and considering only leading terms in $\varepsilon_0/A_\text{mw}$ for $P_\text{LZ,2}$. The good agreement between the experiment and the simulation shows the validity of this simple dynamical picture and demonstrates an efficient way to increase the dispersive readout signal after manipulation by projecting the coherent state to an even-parity charge state. By comparing Eq.~(\ref{CQ_TLS}) and Eq.~(\ref{CQ_enhanced2}), we note the enhancement of the dispersive signal is determined by the amplification factor $\frac{\alpha_+}{\alpha_-}\frac{(\Gamma_\text{L}-\Gamma_\text{R})}{4f_\text{mw}~}$. Eq.~(\ref{CQ_enhanced2}) also emphasizes the key role of asymmetric relaxation rates in the dispersive detection mechanism. Finally, we observe that by performing a double-passage followed by a projective measurement, we can effectively control the coherence of the system which is now determined by the period of the MW excitation: $T_2\sim f_\text{mw}^{-1}$.

The electron evolution in this double passage regime resembles a standard Mach-Zehnder interferometer (Fig.~\ref{fig:doublepass}(d)). After a double-passage through the anticrossing, the (01) branch of the \lq\lq beam\rq\rq\ is detected via relaxation into the (11) state; indicated by the detector D$_{11}$. This illustration corresponds to the case for $\varepsilon_0>0$, and for $\varepsilon_0<0$, the corresponding picture contains the readout of the (10) \lq\lq beam\rq\rq\, by relaxation into the (00) state. 

\subsection{Single-passage regime}

Finally, we explore the regime of large microwave driving amplitude centred around small detuning, where the LZSM interference pattern arising from passages across the ICT disappears (see Fig.~\ref{fig:lzsm_all}(b), yellow triangle). All four charge states are now involved in the dynamic evolution of the system. In this regime, every passage across the ICT is followed by a projective measurement caused by electron tunnelling from the left dot to the source or drain. Without two consecutive passages with a phase coherent charge state superposition, no interference signal manifests, even though the system is being driven at $f_\text{mw}=21$~GHz, much faster than the decoherence rate of the charge qubit.  The optical analogue of this regime resembles again a standard Mach-Zenhder (see Fig.~\ref{fig:doublepass}(e)). However, in this case an additional detector (D$_{11}$ or D$_{00}$) is placed within one of the branches of the electron \lq\lq beam\rq\rq\ after the first beam splitter. This projective measurement collapses the charge superposition state after the first passage, impeding the formation of an interference pattern.

\section{Outlook and Conclusion}

We have realised a LZSM interference experiment in a CMOS transistor, in which we have observed the single-passage, double-passage and multiple-passage regimes of a single-electron charge qubit by adding progressive stages of projective readout. We have used additional levels arising from the interaction of the qubit with a fermionic reservoir to firstly project the coherent state of the qubit, enhancing the interferometric signal and secondly, suppressing the interference pattern completely even though the driving period is faster than the qubit coherence time. These observation raises possibilities of more sophisticated coherent-control experiments where fast microwave pulses can be used for manipulation, followed by the utilization of a dot-to-lead transition for qubit readout with an enhanced signal. So far, LZSM experiments have been typically studied the region near zero detuning; however we have shown that previously unexplored parameter regimes become visible upon involving the dot-to-sea transitions in the far detuning region. In our study, the gate-based reflectometry technique provides a non-invasive probe of the qubit, one which could further be used to distinguish between adiabatic and non-adiabatic processes~\cite{Mizuta2017}. Our simulations, based on an extension of LZSM theory to encompass the case of a resonator-qubit coupled system, match our data well in each regime. In the future, devices with additional tunability of the level couplings and the relaxation rates could provide access to even more complex interferometry experiments opening a door towards silicon-based quantum optics. For example, split-gate CMOS transistors~\cite{Dupont-Ferrier2013,Betz2015} could enable independent control of the dot occupations allowing to explore the effects of Coulomb interactions and Fermi-Dirac statistics in electron optics, while retaining the scalability of CMOS fabrication.

\section{Acknowledgements}

We thank A. J. Ferguson for discussions. The research leading to these results has been supported by the  European Union's Horizon 2020 research and innovation programme under grant agreement No 688539 (http://mos-quito.eu) and Seventh Framework Programme (FP7/2007-2013) through Grant Agreements No. 318397 (http://www.tolop.eu.) and No. 279781; as well as by the Engineering and Physical Sciences Research Council (EPSRC) through UNDEDD (EP/K025945/1). MFGZ acknowledges support from the Winton Programme for the Physics of Sustainability and Hughes Hall College (University of Cambridge). AC acknowledges support from the EPSRC Doctoral Prize Fellowship. FN was partially supported by the RIKEN iTHES Project, MURI Center for Dynamic Magneto-Optics via the AFOSR Award No. FA9550-14-1-0040, the Japan Society for the Promotion of Science (KAKENHI), the IMPACT program of JST, RIKEN-AIST "Challenge Research" program, JSPS-RFBR grant No 17-52-50023, CREST grant No. JPMJCR1676, and the Sir John Templeton Foundation. SNS was partially supported by the State Fund for Fundamental Research of Ukraine (F66/95-2016).

%


\widetext
\clearpage
\begin{center}
	\Large\textbf{Supplementary information for ``A silicon-based single-electron interferometer coupled to a fermionic sea''}
\end{center}
\setcounter{equation}{0}
\setcounter{section}{0}

\setcounter{figure}{0}
\setcounter{table}{0}
\setcounter{page}{1}
\renewcommand{\thefigure}{S\arabic{figure}}

\section{Magneto-spectroscopy and determination of gate couplings}

\begin{figure}[ht!]%
	\includegraphics[scale=0.4]{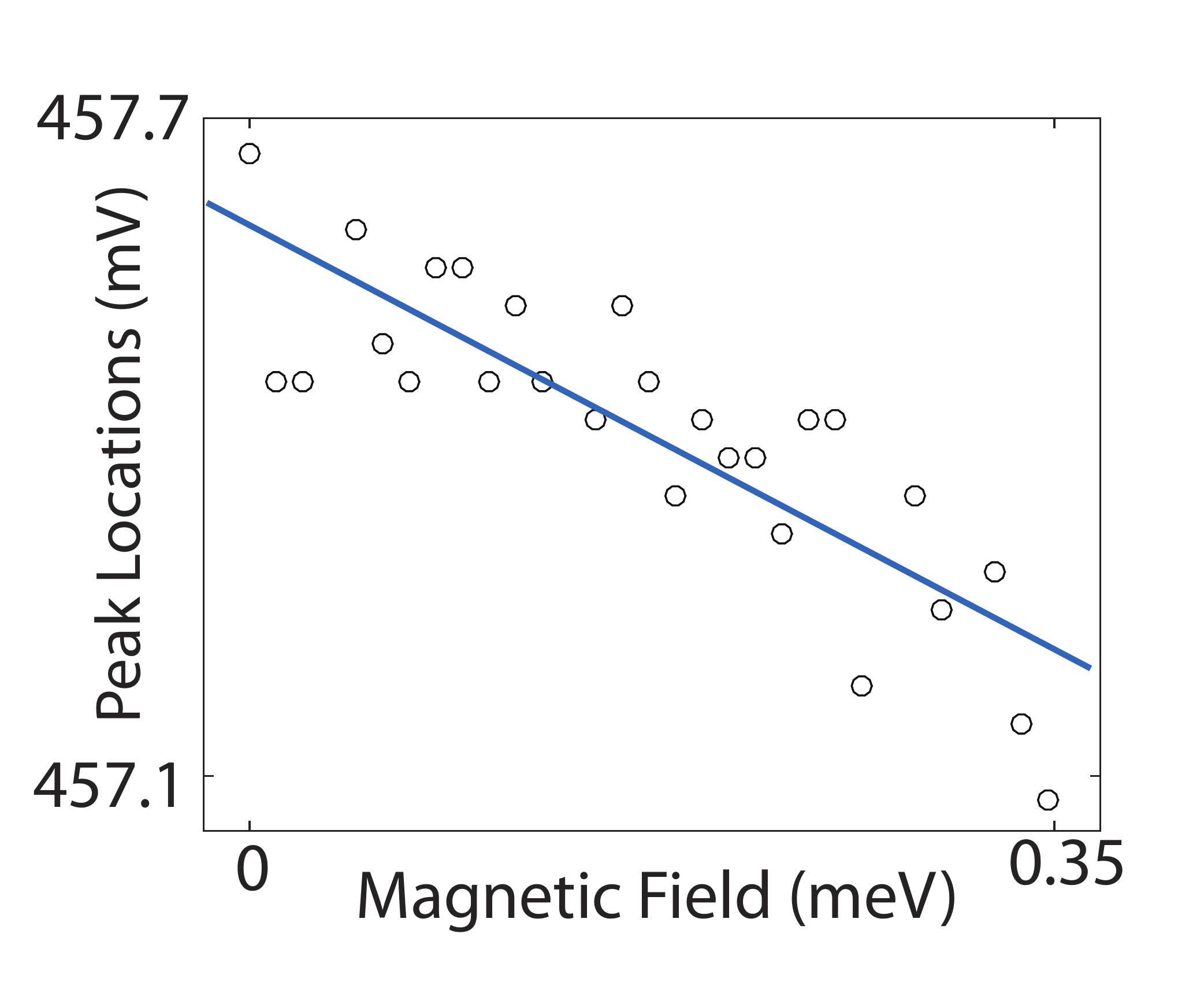}%
	\caption{Peak position in top-gate voltage plotted against the magnetic field energy [assuming $g=2$ (black hollow circles)] and linear fit to the data (blue solid line).
	}
	\label{fig:alpha}%
\end{figure}

In this section, we discuss the calculation of the gate couplings $\alpha_\text{L}$ and $\alpha_\text{R}$ defined as the ratio between the gate capacitance $C_\text{g1(2)}$ and the total capacitance $C_{\Sigma 1(2)}$ for the left (right) dot. This can be done by looking at an absolute energy scale of the system, for example, the Zeeman splitting. In our case, we use magneto-spectroscopy~\cite{Gonzalez-Zalba2014}. We monitor the position in top-gate voltage of the (10)-(11) dot-to-sea transition (DST) shown in Fig.~1(c) in the main text, while sweeping the magnetic field [Fig.~\ref{fig:alpha}]. The shift in the peak location, $\Delta V_\text{TG}$, towards a lower top-gate voltage, as a function of the magnetic field $B$, gives a calibration of the top-gate coupling to the right dot, $\alpha_\text{R}=g\mu_\text{B}B/\Delta V_\text{TG}=0.86$, where $g$ is the electron g-factor and $\mu_\text{B}$ is the Bohr magneton. To obtain the top-gate coupling to the left dot, $\alpha_\text{L}$, we use the fact that, in the multi-passage LZSM regime, interference fringes appear at equidistant values of the top-gate voltage given by the equation $\delta V_\text{TG}=hf_\text{mw}/2e\alpha_-$ [as seen in Fig. 3(c) in the main text]. We obtain $\alpha_\text{L}=0.96$. This calculation also enables us to calibrate to energy the $\varepsilon_\text{0}$ axis in Fig.~2(b), for example. We also calibrate the microwave source voltage output $V_\text{mw}$, to microwave energy amplitude $A_\text{mw}$, using the condition that, at the edge of the LZSM interference region in Fig.~3(c) in the main text, $A_\text{mw}=\varepsilon_\text{0}$.


\section{Calculation of the energy spectrum}

\begin{figure}[ht!]%
	\includegraphics[scale=1]{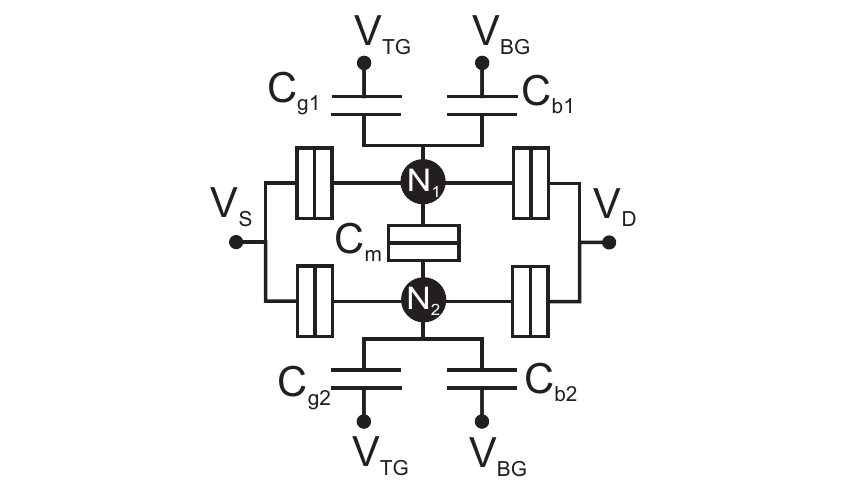}%
	\caption{Circuit diagram for a double quantum dot system, with gate couplings modelled as capacitances. Here, $N_1$ and $N_2$ indicate the electron number in each dot, respectively. The applied voltages are indicated with a $V$, with an appropriate subscript.
	}
	\label{fig:circuit}%
\end{figure}

We calculate the energy spectrum of the double quantum dot system $\Delta E(N_1,N_2)$, where $N_1$ and $N_2$ are the charges in the left and right dot respectively, extending the results from Ref.~\cite{Wiel2002}

\begin{equation}\label{DQD}
\begin{aligned}
\Delta E(N_1,N_2) &= \frac{1}{2}N_1^2E_\text{C1}+\frac{1}{2}N_2^2E_\text{C2}+N_1N_2E_\text{Cm}\\
&-\frac{1}{\left|e\right|}\left\{C_\text{g1}V_\text{TG}(N_1E_\text{C1}+N_2E_\text{Cm})+C_\text{g2}V_\text{TG}(N_1E_\text{Cm}+N_2E_\text{C2})\right.\\
&\left.+C_\text{b1}V_\text{BG}(N_1E_\text{C1}+N_2E_\text{Cm})+C_\text{b2}V_\text{BG}(N_1E_\text{Cm}+N_2E_\text{C2})\right\}\\
&+\frac{1}{e^2}\left\{\frac{1}{2}C_\text{g1}^2V_\text{g}^2E_\text{C1}+\frac{1}{2}C_\text{g2}^2V_\text{g}^2E_\text{C2}+C_\text{g1}C_\text{g2}V_\text{g}^2E_\text{Cm}\right.\\
&\left.+\frac{1}{2}C_\text{b1}^2V_\text{BG}^2E_\text{C1}+\frac{1}{2}C_\text{b2}^2V_\text{BG}^2E_\text{C2}+C_\text{b1}C_\text{b2}V_\text{BG}^2E_\text{Cm}\right\}.
\end{aligned}
\end{equation}

\noindent where the various parameters of the double-dot system are described by the simplified circuit diagram shown in Fig.~\ref{fig:circuit}. In Eq.~\ref{DQD}, $E_\text{C1}$, $E_\text{C1}$ and $E_\text{Cm}$ correspond to the charging energy of the left dot, right dot and mutual energy, respectively.

%
%
%

In order to simplify the expression, we define several new variables. We introduce the reduced top-gate and back-gate voltages, $n_\text{t}$ and $n_\text{b}$, respectively,

\begin{equation}
\begin{aligned}
n_\text{t}&=&\frac{C_\text{g1}V_\text{TG}}{\left|e\right|}\\
n_\text{b}&=&\frac{C_\text{b1}V_\text{BG}}{\left|e\right|}.\\
\end{aligned}
\end{equation}

Additionally, we consider the following approximate charging energy relation

\begin{equation}
E_\text{C}=E_\text{C1}=E_\text{C2}=mE_\text{Cm},
\end{equation}

\noindent where {\itshape m} accounts for the ratio between the charging energy and mutual energy. Moreover, we may introduce the asymmetry in the gate couplings {\itshape a} defined by:

\begin{equation}
a=\frac{2\left|\alpha_-\right|}{\text{max}(\alpha_\text{L},\alpha_\text{R})}.
\end{equation}

\noindent where $\alpha_-=(\alpha_\text{L}-\alpha_\text{R})/2$. We can now express the ratio of the DQD energies to the total charging energy as

\begin{equation}
\begin{aligned}
\frac{\Delta E(N_1,N_2)}{E_C} &= \frac{1}{2}N_1^2+\frac{1}{2}N_2^2+\frac{N_1N_2}{m}\\
&-n_t\left\{N_1+\frac{N_2}{m}+(1+a)\left(\frac{N_1}{m}+N_2\right)\right\}\\
&-n_b\left\{N_1+\frac{N_2}{m}+(1-a)\left(\frac{N_1}{m}+N_2\right)\right\}\\
&-n_t^2\left\{\frac{1}{2}+\frac{1}{2}(1+a)^2+\frac{1+a}{m}\right\}\\
&-n_b^2\left\{\frac{1}{2}+\frac{1}{2}(1-a)^2+\frac{1-a}{m}\right\}.\\
\end{aligned}
\end{equation}

We find that the reduced back-gate voltage value at which we perform the experiment -- the middle of the ICT line -- corresponds to $n_\text{b}^0=0.25$. Finally, we calculate the reduced-energy diagram of the DQD across the $n_\text{b}=0.25$ line, as shown in Fig.~2(a) in the main text. We do so by plotting $\Delta E/E_\text{C}(N_1,N_2)$ for $N_1,N_2=0,1$ as a function of reduced detuning $\varepsilon_0/E_\text{C}=2\alpha_-n_\text{t}/\alpha_\text{L}$, for $m=10$ and $a=0.1$, as in the case of the experiment. We use a tunnel coupling $\Delta_\text{c}=E_\text{C}/150$.

%
%
%


\section{Frequency dependence of the LZSM interference pattern}

\begin{figure}[ht!]%
	\includegraphics[scale=0.4]{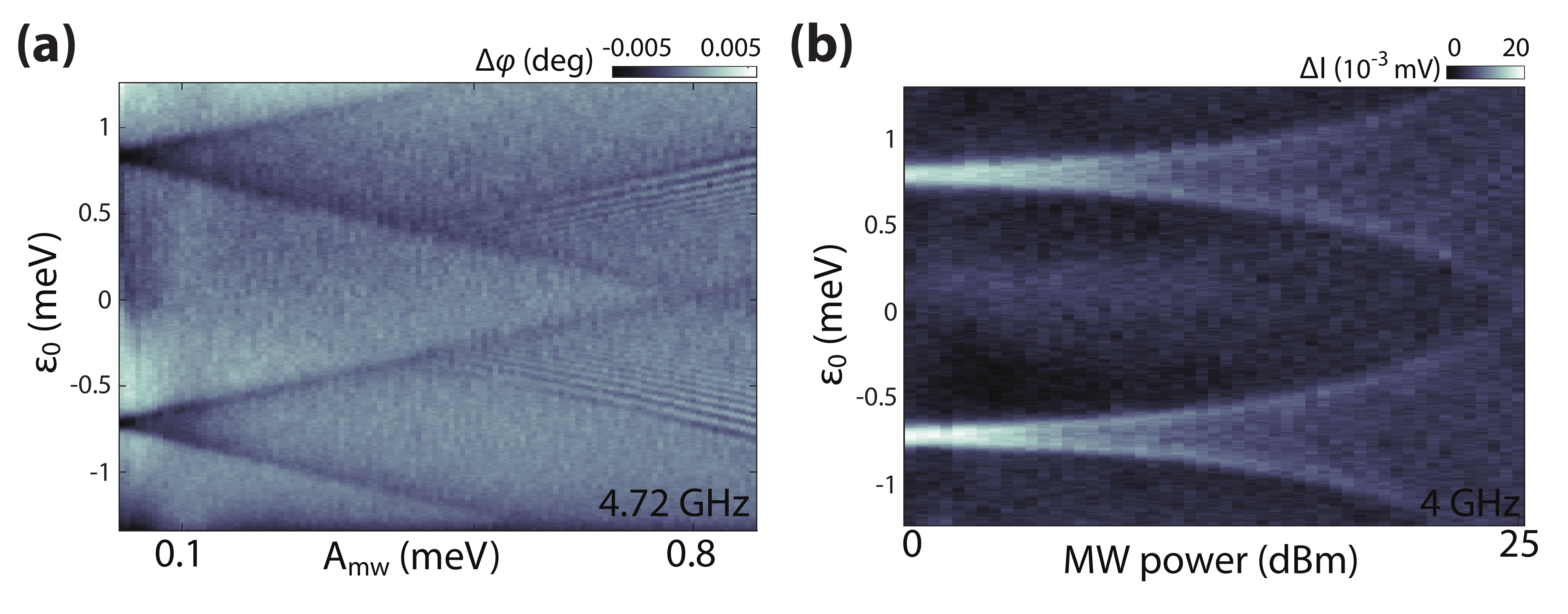}%
	\caption{(a) Resonator phase response as a function of detuning $\epsilon_\text{0}$ and the microwave amplitude at (a) 4.72\,GHz (b) In-phase component of the resonator response at 4\,GHz, showing the decline and disappearance of the LZSM interference pattern.
	}
	\label{fig:freqdep}%
\end{figure}

The microwave frequency used to drive the LZSM experiment can have a significant effect on the interference pattern, depending on the qubit timescales. If the microwave frequency is lowered below the electron phase coherence time $T_\mathrm{2}$, the system goes into the incoherent regime, where the electron phase coherence is lost before two consecutive passages are performed. In our double quantum dot, $T_\mathrm{2}$ can therefore be studied by performing a microwave-frequency dependence of the LZSM interferometry pattern.  Figure \,\ref{fig:freqdep} shows the LZSM interferometry results obtained using two different microwave frequencies ($4.72$\,GHz and $4$\,GHz). In (a), taken at $4.72$\,GHz, we observe a few lines of interference in the double-passage region, indicating that $f_\text{mw}^{-1}$ is approaching $T_\mathrm{2}$. In (b), taken at $4$\,GHz, there are almost no visible interference fringes indicating $f_\text{mw}^{-1}\simeq T_\mathrm{2}$. Below $4$\,GHz the interference pattern disappears, indicating that we have an upper limit of $T_2\simeq 250$\,ps.
\section{Stuckelberg phase for the double-passage experiment}

The St{\"u}ckelberg phase, $\phi_\text{St}$, that encapsulates the phase difference acquired after the first passage, can be derived in the adiabatic-impulse model by considering a driven two-level system~\cite{Shevchenko2010}. In the fast-passage limit $(\Delta_\text{c}^{2}/A_\text{mw}\omega <1)$, the upper-level probability after two passages reads,

\begin{eqnarray}
P_{\mathrm{doub}} &\simeq &2\pi \frac{\Delta_\text{c} ^{2}}{A_\text{mw}\omega }\left( 1-\frac{%
	\varepsilon _{0}^{2}}{A_\text{mw}^{2}}\right) ^{-1/2}\sin ^{2}\phi _{\mathrm{St}},
\label{PII_1} \\
\phi _{\mathrm{St}} &=&-\frac{\varepsilon _{0}}{\omega }\arccos\left(\frac{%
	\varepsilon _{0}}{A_\text{mw}}\right)+\frac{A_\text{mw}}{\omega }\left( 1-\frac{\varepsilon _{0}^{2}}{%
	A_\text{mw}^{2}}\right) ^{1/2}\!\!\!-\frac{\pi }{4}  \label{FiSt}
\end{eqnarray}%
\noindent where $\omega= 2\pi f_\text{mw}$. Equation~(\ref{PII_1}) can be simplified in leading order with $\varepsilon _{0}/A_\text{mw}$:%
\begin{equation}
P_{\mathrm{double}}\simeq \frac{2\pi }{\omega }\frac{\Delta_\text{c} ^{2}}{A_\text{mw}}\sin ^{2}%
\left[ \frac{A_\text{mw}}{\omega }-\frac{\pi }{2}\frac{\varepsilon _{0}}{\omega }-%
\frac{\pi }{4}\right] .  \label{PII_2}
\end{equation}

Finally, we use this expression to calculate the dispersive response of the resonator in Fig.~4(c) in the main text. We note that both the amplitude and frequency dependence in Eq.~(\ref{PII_2}) is in agreement with the result of Ref.~\cite{Berns2006} in the linear approximation. And this didactic result comes as a nice surprise, since the latter result was obtained in the assumption of small $\Delta_\text{c}$ and fast and strong driving, while the former result was obtained in the LZSM picture of adiabatic-impulse model, with impulse-type transitions between adiabatic energy levels.

\section{Incoherent regime}

\begin{figure}
	\centering
	\includegraphics{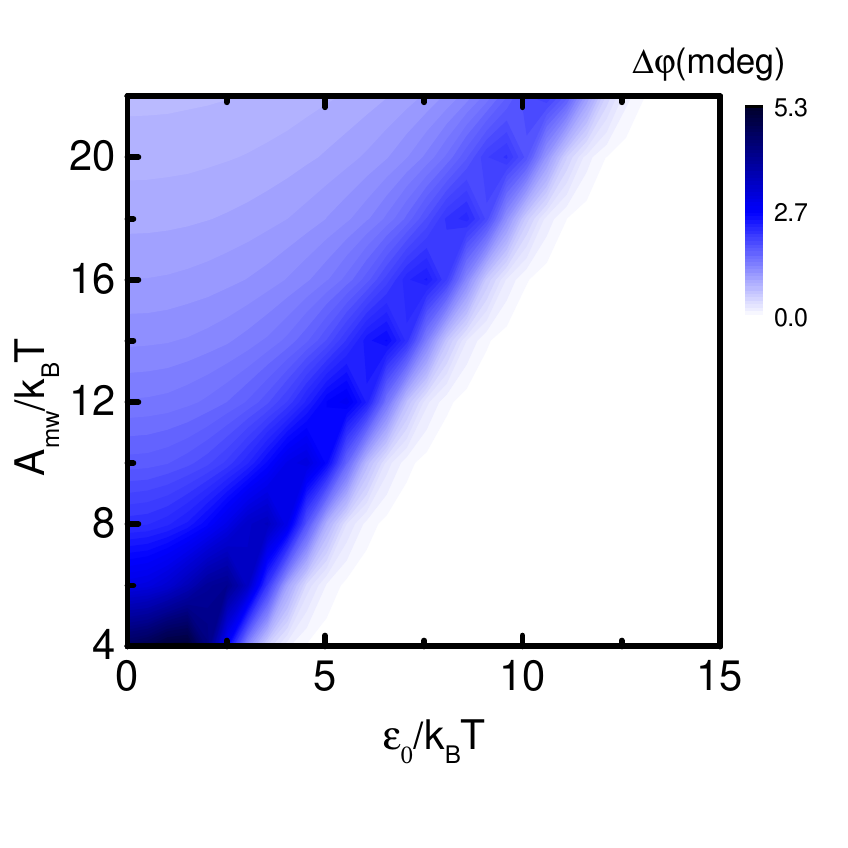}
	\caption{Resonator phase response $\Delta\varphi$ as a function of reduced detuning $\varepsilon_0/k_\text{B}T$ and reduced microwave amplitude $A_\text{mw}/k_\text{B}T$}
	\label{fig:incoherent}
\end{figure}

In this section, we calculate the resonator response $\Delta\varphi$ in the incoherent regime, when single-electrons are exchanged at microwave frequencies between the left (or right) QD and the source and drain reservoirs. This scenario can be accurately described by a fast-driven two-level system~\cite{Persson2010,Gonzalez-Zalba2015}. In this situation, we calculate the parametric capacitance of a single QD, given by the expression,

\begin{equation}
C_\text{pm}=-\left(e\alpha_\text{L}\right)^2\frac{\partial \overline{P_1}(t)}{\partial\varepsilon_0}.
\end{equation}

\noindent where $\overline{P_1}$ is the average probability of having one excess electron in the left QD. Also, $\overline{P_1}$ can be calculated from the time-dependent expression of the probability subject to a sinusoidal change in energy detuning  $\varepsilon(t)=\varepsilon_0+A_\text{mw}\text{sin}(2\pi f_\text{mw}t)$ induced by a microwave electric field with amplitude $A_\text{mw}$ and frequency $f_\text{mw}$ around an energy detuning offset $\varepsilon_0$,

\begin{equation}
P_1(t)\simeq\frac{1}{1+\text{exp}\left(\frac{\varepsilon(t)}{k_\text{B}T}\right)}
\end{equation}

\noindent where $k_\text{B}$ is the Boltzmann constant and $T$ the electron temperature. In Fig~\ref{fig:incoherent}, we present the resonator phase response as a function of $\varepsilon_0$ and $A_\text{mw}$, for a resonator quality factor $Q=42$, $\alpha_\text{L}=0.96$, and $T=100$~mK. The calculation matches well the data in Fig. 2(b) (blue-star regions). Particularly, it captures the enhancement of the phase signal at $\varepsilon_0=A_\text{mw}$.

\section{Relaxation rates for the dot-to-lead transitions}

In this section, we provide a quantitative analysis of the relaxation rates between the left (right) dot and the electron reservoirs, $\Gamma_\text{L}$ $(\Gamma_\text{R})$.  As discussed in the main text, the difference in these relaxation rates results in relaxation occurring primarily at the (01)-(11) and (00)-(10) DSTs rather than at the (00)-(01) and (10)-(11) transitions when operating the system in the strong driving regime. This can be directly observed in Fig.~2(b) of the main text, where no additional observable change in $\Delta\varphi$ appears in the regions $A_\text{mw}<\left|\varepsilon_0-\varepsilon^{00-01}_0\right|$ and $A_\text{mw}<\left|\varepsilon_0-\varepsilon^{10-11}_0\right|$, where $\varepsilon^{00-01}_0$ and $\varepsilon^{10-11}_0$ correspond to the position in detuning of the (00)-(01) and (10)-(11) crossings, respectively. In the following, we provide additional experimental evidence demonstrating that $\Gamma_\text{L}\gg\Gamma_\text{R}$.

First, we extract $\Gamma_\text{L}$. For the (01)-(11) DST, the probability of an electron in the left dot to relax into the source-drain reservoir is given by $P_\text{SD}=1-\exp\left(-\Gamma_\text{L}\Delta t\right)$, where the $\Delta t$ is the time the electron spends at a value of detuning larger than the position of the (01)-(11) crossing ($\varepsilon^{01-11}_0$). It can be shown that

\begin{equation}
\Delta t=\frac{\pi}{\omega}-\frac{2}{\omega}\sin^{-1}\left(\frac{\varepsilon^{01-11}_0-\varepsilon_0}{A_\text{mw}}\right)=\frac{1}{\omega}\left[\pi-2\sin^{-1}(\frac{\varepsilon^{01-11}_0-\varepsilon_0}{A_\text{mw}})\right],
\end{equation}
which then gives
\begin{equation}
P_\text{SD}=1-\exp\left\{\frac{-\Gamma_\text{L}}{\omega}\left[\pi-2\sin^{-1}\left(\frac{\varepsilon^{01-11}_0-\varepsilon_0}{A_\text{mw}}\right)\right]\right\}.
\label{eq:gammal}
\end{equation}

\begin{figure}[ht!]%
	\includegraphics[scale=0.3]{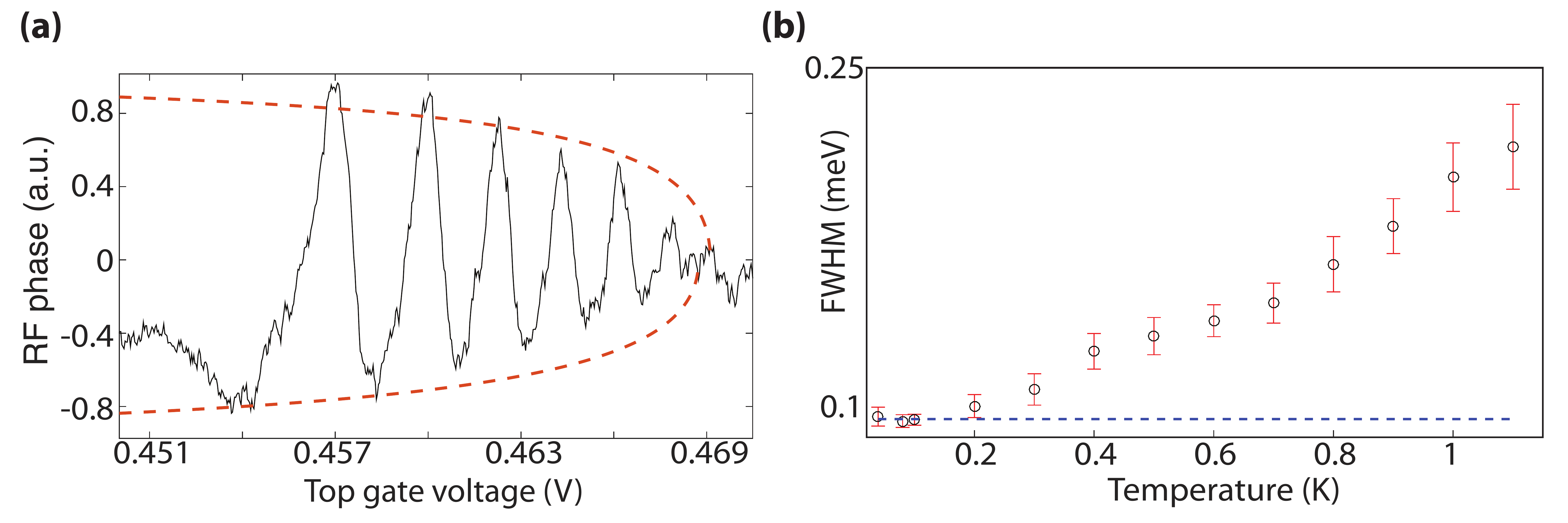}%
	\caption{ (a) Resonator phase response as a function of top gate voltage V$_\text{TG}$, with a fit to the envelope of the oscillations as given by Eq.\,\ref{eq:gammal}. (b) Full width at half maximum of the DST peak with varying mixing chamber temperature. The blue dashed line shows the saturation of the peak FWHM at low temperature.
	}
	\label{fig:relaxation}%
\end{figure}

Since in the double-passage regime, the probability $P_{11}$ is proportional to $P_\text{SD}$, we fit Eq.~(\ref{eq:gammal}) to the envelope of the oscillations (which decay along detuning axis as predicted by the equation), as shown in Fig.\,\ref{fig:relaxation}(a), for a trace taken at $f_\text{mw}=15$\,GHz. The fit gives a tunnel rate of $\Gamma_\text{L}\approx50$\,GHz.

To extract $\Gamma_\text{R}$, we perform an analysis of the temperature dependence of (10)-(11) DTS. In Fig.~\ref{fig:relaxation}, we show the FWHM of the $\Delta\varphi$ peak as a function of the temperature of the mixing chamber. As we lower the temperature, we see a linear decrease of the FWHM until 700~mK. Decreasing the temperature further leads to a saturation of the FHWM to 100~$\mu$eV below 200~mK. The mechanism that can lead to saturation of the FWHM can be originated by either electron-phonon decoupling or lifetime broadening~\cite{House2015}. In the case of electron-phonon decoupling, the saturation occurs at FWHM$=3.5k_\text{B}T$ and in the case of lifetime-broadening FWHM$=2h\Gamma_\text{R}$. From the data, we are unable to distinguish the exact mechanism that causes saturation but we can extract an upper bound for $\Gamma_\text{R}<12$~GHz.

\end{document}